\documentclass[%
 reprint,
superscriptaddress,
 amsmath,amssymb,
 aps,
 longbibliography,
floatfix
]{revtex4-1}

\usepackage{graphicx}
\usepackage{dcolumn}
\usepackage{bm}

\begin{document}

\title{Single spin sensing of domain wall structure and dynamics in a thin film skyrmion host}

\author{Alec Jenkins}
\affiliation{%
 Department of Physics, University of California, Santa Barbara, Santa Barbara, California 93106, USA\\
}%
\author{Matthew Pelliccione}%
\affiliation{%
 Department of Physics, University of California, Santa Barbara, Santa Barbara, California 93106, USA\\
}%
\author{Guoqiang Yu}
\affiliation{
 Beijing National Laboratory for Condensed Matter Physics, Institute of Physics, Chinese Academy of Sciences, Beijing 100190, China
}%
\author{Xin Ma}
\affiliation{%
 Department of Electrical and Computer Engineering, University of California, Santa Barbara, Santa Barbara, California 93106, United States\\
}%
\author{Xiaoqin Li}
\affiliation{%
 Department of Physics, The University of Texas at Austin, Austin, Texas 78712, United States\\
}%
\author{Kang L. Wang}
\affiliation{
 Department of Electrical Engineering, University of California, Los Angeles, California 90095, United States
}%
\author{Ania C. Bleszynski Jayich}%
\affiliation{%
 Department of Physics, University of California, Santa Barbara, Santa Barbara, California 93106, USA\\
}%

\date{\today}

\begin{abstract}
Skyrmions are nanoscale magnetic structures with features promising for future low-power memory or logic devices. In this work, we demonstrate novel scanning techniques based on nitrogen vacancy center magnetometry that simultaneously probe both the magnetic dynamics and structure of room temperature skyrmion bubbles in a thin film system Ta/CoFeB/MgO. We confirm the handedness of the Dzyaloshinskii-Moriya interaction in this material and extract the helicity angle of the skyrmion bubbles. Our measurements also show that the skyrmion bubbles in this material change size in discrete steps, dependent on the local pinning environment, with their average size determined dynamically as their domain walls hop between pinning sites. In addition, an increase in magnetic field noise is observed near all skyrmion bubble domain walls. These measurements highlight the importance of interactions between internal degrees of freedom of skyrmion bubble domain walls and pinning sites in thin film systems. Our observations have relevance for future devices based on skyrmion bubbles where pinning interactions will determine important aspects of current-driven motion.
\end{abstract}

\maketitle

\section{\label{sec:Introduction}Introduction}

Magnetic skyrmions are solitonic spin textures with non-trivial topology. They were first discovered in non-centrosymmetric bulk crystals and ultrathin epitaxial magnetic layers \cite{Muhlbauer2009, Yu2011, Heinze2011, Schulz2012}. The skyrmions observed in those materials had nanoscale sizes and large current-driven velocities, leading to their identification as promising candidates as carriers of information in future high-density, low-power electronics \cite{Jonietz2011, Sampaio2013, Fert2013, Yu2012}. Recently, skyrmions have been observed at room temperature in sputtered thin film multilayers \cite{Woo2016, Yu2016, Moreau-Luchaire, Soumyanarayanan2017}. These multilayer materials are promising for making practical skyrmion devices, but further materials development is required to achieve the 10 nm-scale and efficient current-driven motion required by applications \cite{Fert2017}.

\begin{figure*}[t]
\includegraphics[width=\textwidth]{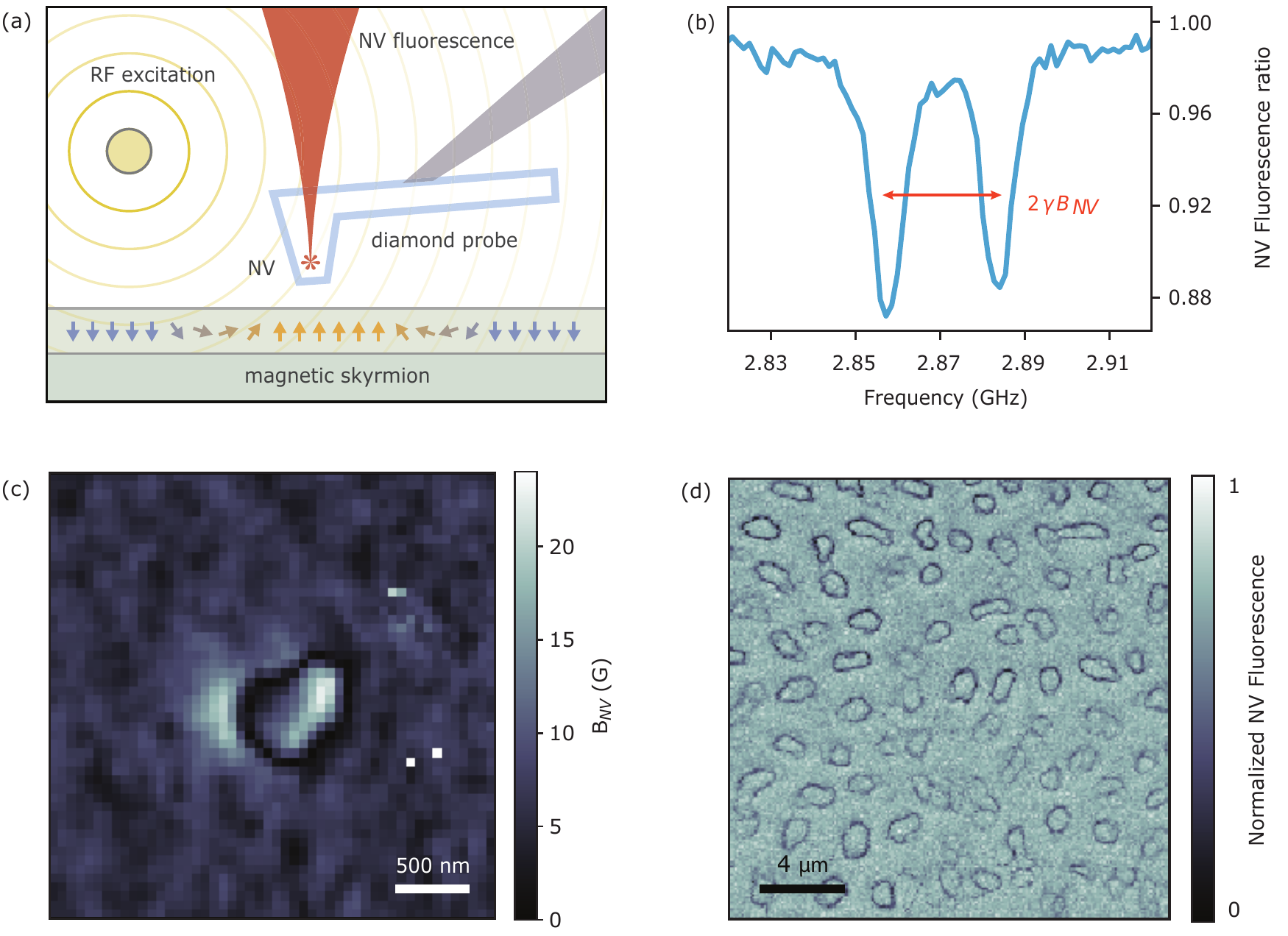}
\caption{ Scanning NV-based imaging of skyrmions. (a) Diagram of the experimental setup. A single-crystal diamond probe containing a single NV center near the apex of its tip is scanned above the multilayer sample. Simultaneous optical and RF excitation of the NV gives an ESR signal (b), which is used to measure the stray magnetic field at each position in the scan. (b) Example of an NV ESR signal: the NV fluorescence rate decreases when applied microwaves are on resonance with either of the NV’s two spin transitions ($m_s = 0 \rightarrow +1$ and $m_s = 0 \rightarrow -1$). The splitting of the two peaks is used to calculate $B_{NV}$, the magnetic field along the NV center axis. Plotted is the ratio of the NV fluorescence to its off-resonant value. (c) NV magnetic image of a skyrmion bubble with an external magnetic field of 10.0 Oe perpendicular to the film plane. (d) Magnetic contour image with applied microwave frequency = 2.870 GHz and with 6.5 Oe perpendicular to the film plane.
}
\label{fig1}
\end{figure*}

As skyrmion multilayers and devices are developed, local, real-space probes of these systems will be crucial to understanding their microscopic behavior. Measurements of skyrmion size and structure, current-driven behavior, and interaction with defects will be necessary to engineer materials with characteristics optimized for use in skyrmion devices. Several imaging techniques have already been used to study skyrmions, including Kerr microscopy \cite{Yu2016, Jiang}, Lorentz transmission electron microscopy \cite{Yu2010a, Pollard2017}, transmission X-ray microscopy \cite{Woo2016}, and magnetic force microscopy \cite{Soumyanarayanan2017}. While each of these tools has certain advantages, a scanning probe microscope (SPM) based on the nitrogen-vacancy (NV) defect in diamond \cite{Taylor2008, Degen2008, Balasubramanian2008} is another probe particularly well-suited to studying skyrmion devices; the NV-SPM is magnetically non-invasive and offers quantitative, nanoscale spatial resolution of stray magnetic fields over a wide range of temperatures and magnetic fields \cite{Maletinsky2012, Rondin2013, Dovzhenko2018, Gross2016, Pelliccione2016, Gross2018}. 

In this work, we utilize a scanning NV microscope to study magnetic skyrmions in a multilayer system Ta(2nm)/Co$_{20}$Fe$_{60}$B$_{20}$(1nm)/Pt(1\AA)/MgO(2nm) \cite{Yu2016}. We demonstrate that scanning NV microscopy is a uniquely versatile tool for the study of skyrmion materials, useful for investigating both magnetic structure and dynamics in these materials. Specifically, we use the NV microscope to extract magnetic parameters of this thin film system such as the exchange stiffness and the domain wall width and then employ these parameters to study the chiral nature of the skyrmion spin structure. Importantly, our measurements show ubiquitous interactions between pinning sites and the internal degrees of freedom of skyrmion bubble domain walls. Pinning of skyrmion internal degrees of freedom has previously been shown to be a key factor in determining skyrmion size in similar multilayer systems \cite{Gross2018}. Here, we observe that skymrion bubble sizes change in discrete steps as a function of the applied external field. For many of the observed skyrmion bubbles, their average size is determined dynamically as sections of the bubble domain walls hop back and forth between multiple pinning sites. We quantitatively probe the dynamics of these hopping processes using spectroscopy of an NV center positioned near fluctuating bubble walls. We also image the local magnetic noise environment near skyrmion bubbles and observe a universal increase in magnetic noise near skyrmion bubble domain walls. Our direct observations of skyrmion bubbles interacting with multiple pinning sites have important implications for skyrmion bubble motion in this material. As shown previously, high densities of pinning sites can lead to a break down of typical micromagnetic models of skyrmion motion, which show low depinning current densities and large velocities in the presence of sparse defects \cite{Woo2016}.

\section{\label{sec:Experiment}Experimental setup}

\begin{figure*}[t]
\includegraphics[width=\textwidth]{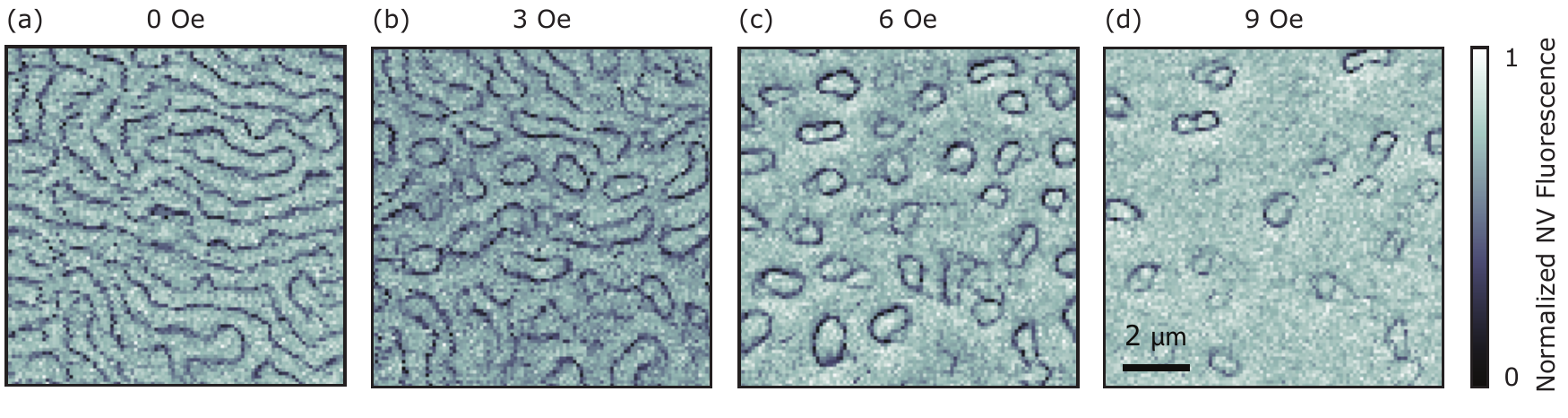}
\caption{\label{fig2} Field contour images of magnetic phases in Ta/CoFeB/MgO system acquired with the NV microscope. The applied microwave frequency is 2.870 GHz for all images. An external magnetic field applied normal to the sample plane is increased from 0-9 Oe in (a)-(d). As the field increases, the magnetic order evolves from stripes at 0 Oe (a) into a mixed skyrmion/stripe phase at 3 Oe (b) and a skyrmion phase at 6 (c) and 9 (d) Oe. At these fields, zero-field contours mark the approximate domain wall positions. The scale bar for all images is shown in (d).
}
\end{figure*}

In these measurements, a single NV center in a diamond probe is scanned over the sample surface while measuring stray magnetic fields using the NV center's optically detected magnetic resonance spectrum (Fig. 1), hence referred to as its (electron spin resonance) ESR spectrum \cite{Pelliccione2016}. This spectrum is measured by sweeping the frequency of applied microwaves while monitoring the spin-state-dependent NV fluorescence rate (Fig. 1b). Magnetic fields induce a Zeeman splitting $\Delta f$ between the NV $m_s = \pm 1$ spin states that is proportional to the absolute value of the stray field along the axis of the NV--- $\Delta f \simeq 2\gamma |B_{NV}|$, where $\gamma = 2.8$ MHz/G is the gyromagnetic ratio of the NV. $B_{NV}$ is calculated for a given ESR spectrum using this Zeeman splitting with a small correction due to fields perpendicular to the NV axis \cite{VanderSar2015}. ESR measurements are used to acquire a two-dimensional map of $B_{NV}$ (Fig. 1c). Based on this map, it is possible to reconstruct all vector components of the stray field \cite{Dreyer2007, Lima2009}, and the reconstructed vector field can be used to estimate the domain wall positions comprising an individual magnetic bubble and probe the internal structure of domain walls \cite{Tetienne2015, Dovzhenko2018, Gross2016}. The imaging resolution of this technique is determined by the distance from the NV to the magnetic CoFeB layer. For the image in Fig. 1c, this distance was measured to be 58 $\pm$ 5 nm (supplementary section S2). 

Measuring the ESR spectrum at each scan point is time intensive, so a faster contour imaging method is used to get information about magnetic structure. In a contour measurement, the frequency of applied microwaves is fixed while the microwave amplitude is square-wave modulated on/off at kHz frequencies. The NV fluorescence rate difference for microwaves on vs.\ off is measured at each point of the scan. Dark contours in the resulting image correspond to resonances of the applied microwaves with the $m_s = 0$ to $m_s \pm 1$ transitions--- to first order giving contours of constant magnetic field. These contours, with a properly chosen microwave frequency, correspond to domain wall locations in the underlying thin film, as discussed below. The contour image in Fig. 1d outlines the approximate domain wall positions of a group of skyrmion bubbles.

\section{\label{sec:phases}Magnetic phases T\lowercase{a}/C\lowercase{o}F\lowercase{e}B/M\lowercase{g}O}

In the Ta/CoFeB/MgO thin film system, perpendicular magnetic anisotropy (PMA) due to the CoFeB/MgO interface allows for the existence of magnetic bubbles stabilized under a small magnetic field perpendicular to the film. An interfacial Dzyaloshinskii-Moriya interaction (DMI) arising from an antisymmetric exchange coupling at the interface of the ferromagnetic CoFeB and strong spin-orbit Ta layers encourages a fixed chirality of the magnetization structure of these bubbles. A wedged sub-nm Pt insertion layer between the CoFeB and MgO tunes the PMA strength by weakening Co-O and Fe-O bonds at the MgO interface. This Pt insertion layer also induces a DMI at the top CoFeB interface, modifying the total DMI strength \cite{Ma2016, Ma2017, Ma2018}. A magnetic field $B_{ext}$ is applied perpendicular to the sample plane giving rise to isolated skyrmions in certain regions of the PMA-$B_{ext}$ phase space \cite{Yu2016}. The NV fluorescence images in Fig. 2 show the evolution of the magnetic order with $B_{ext}$. At $B_{ext} = 0$, magnetic order takes the form of stripe-like domains. As $B_{ext}$ is increased, the system undergoes a first order phase transition into a skyrmion phase. In Fig. 2, a coexistence of skyrmion bubbles and stripes is seen already at 3 Oe, and several bubbles persist up to $B_{ext}=9$ Oe. At fields larger than 10 Oe, the material transitions into a ferromagnetic phase with no magnetic features in the NV images. The images in Fig. 2 were obtained using the contour imaging method, with the frequency of the applied microwaves fixed to the NV zero-field splitting frequency of 2.870 GHz. For the small values of $B_{ext}$ used, the dark contours mark the approximate domain wall positions, with a small ($<$ 100 nm) scale offset in the direction of the NV's in-plane projection (supplementary section S1). The width of the contour lines, which can be much smaller than the NV-sample separation, is determined by the width of the NV ESR dip and the magnetic field gradients near the domain walls \cite{Myers2014a}.

\section{\label{sec:dynamics}Skyrmion bubble pinning dynamics}

\begin{figure*}[t]
\includegraphics[width=\textwidth]{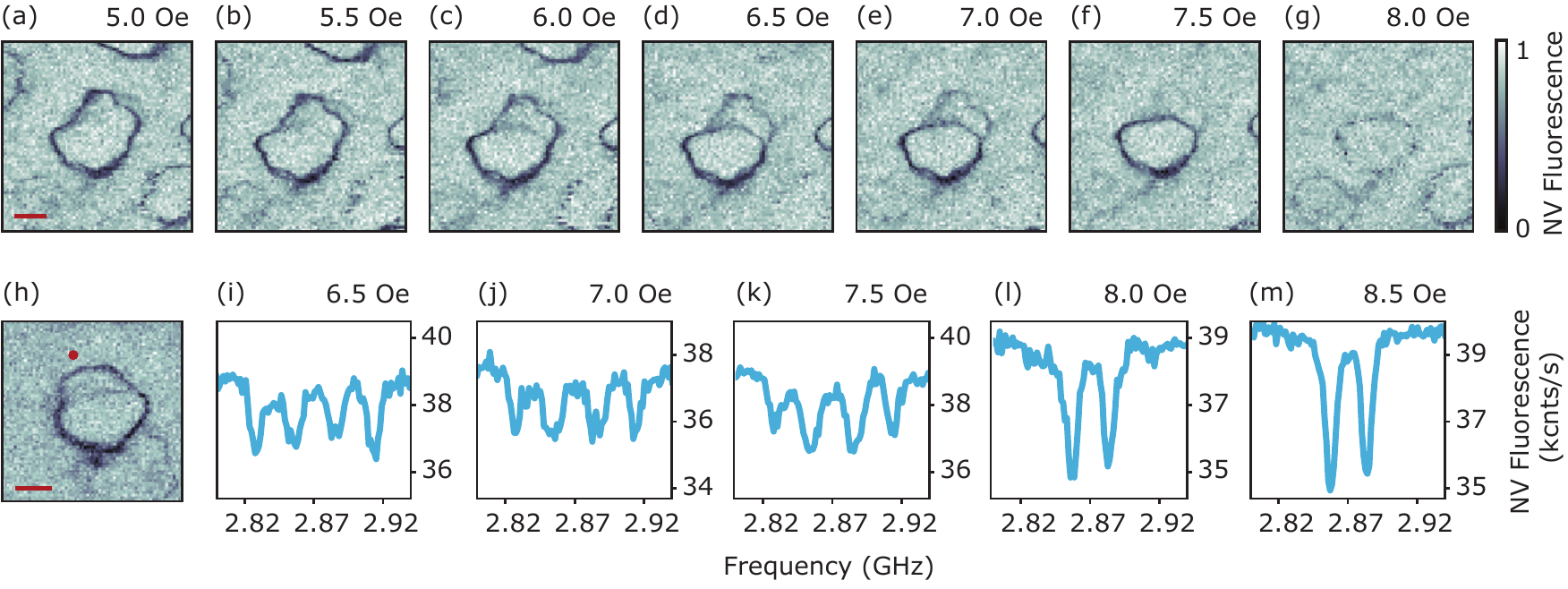}
\caption{\label{fig3} Domain wall fluctuations and evolution with $B_{ext}$ imaged by NV center microscopy. (a) 2.870 GHz contour images of a skyrmion bubble, each labeled by the corresponding $B_{ext}$ applied during that scan. As $B_{ext}$ increases, a section of the domain wall evolves into a bistable configuration, seen most clearly in the (d) and (e) scans with 6.5 and 7.0 Oe. The scale bar for (a)-(g) is shown in (a). (h) 2.870 GHz contour images of another skyrmion bubble taken at $B_{ext} = 7.0$ G. (i)-(m) NV center ESR spectra taken with the NV fixed at the location indicated by the red dot, each labeled by $B_{ext}$. The mulitple pairs of ESR splittings in each spectrum correspond to different positions of the domain wall.
}
\end{figure*}

The high resolution of the NV center scanning microscope allows for the study of the microscopic structure of these bubble domains \cite{Gross2016,Dovzhenko2018}. For example, contour images of domain wall position show the effects of pinning sites on skyrmion shape (Fig. 3a), which induce both a static deformation of the skyrmion as well as a dynamic instability. Figures 2c-d and the higher resolution contour images in Fig. 3a-g show irregularly shaped skyrmions, whose dramatic deviation from a disorder-free, circular shape is consistent with previously reported NV-microscopy images of a similar thin film magnetic multilayer \cite{Gross2018}. The effect of pinning sites is also manifest in the evolution of skyrmion size and shape with magnetic field, as shown in Fig. 3. When $B_{ext}$ is increased from 5 to 7.5 Oe the skyrmion shrinks, as seen by comparing Fig. 3a and 3f and as predicted by micromagnetic theory \cite{Thiele1970}. Interestingly, however, this process does not happen smoothly but rather discontinuously: at intermediate fields in the range of $B_{ext} = $ 6.0--7.5 Oe, the images in Fig. 3b-e show domain wall contours corresponding to both the larger and smaller diameter skyrmion. As the field is increased, the contrast of the larger diameter contour progressively decreases while the contrast of the smaller diameter contour increases. This behavior is explained by the domain wall hopping back and forth in time between two stable positions, progressively spending a larger fraction of its time in a smaller diameter configuration as the field is increased. Hopping that occurs on a timescale faster than the NV measurement leads to a reduction in the contrast of the contours because the NV fluorescence signal is averaged over its bright and dark states as the fluctuating field produced by the hopping domain wall brings the applied microwaves on and off resonance with the NV ESR transitions. Thus, although our measurement is too slow to detect the telegraph nature of the domain wall hopping in real time, we can detect time-averaged signatures of the dynamics through changes in contour contrast. 


We can confirm the time-averaged behavior of the domain wall fluctuations by fixing the NV at a location near a fluctuating domain wall while recording the ESR spectrum, as shown in Fig. 3i-m. The position of the NV is indicated by the red dot in Fig. 3h. In the spectrum, the hopping of the domain wall appears as two dominant ESR splittings that emerge as the magnetic field is swept through the skyrmion phase. When the domain wall is near the NV, the ESR splitting is largest, given by the outer two ESR dips. The existence of other ESR dips in the spectra in Fig. 3 implies that the domain wall spends some time at another position, seen as the faint contour line cutting across the middle of the bubble in Fig. 3h. Qualitatively, the evolution of the contrast ratio between different pairs of dips in the ESR spectrum or equivalently, between domain wall branches in the contour images, gives an indication of the relative time spent in different domain wall states. As $B_{ext}$ is increased, the outer pair of ESR dips grows fainter as the domain wall evolves from spending more time in the larger skyrmion diameter configuration (near the NV) to spending more time in the small diameter configuration. At $B_{ext}=$ 8.5 Oe, the skyrmion bubble is no longer stable and the ESR splitting is given by the NV-axis projection of $B_{ext}$.

\begin{figure}
\includegraphics[width=\columnwidth]{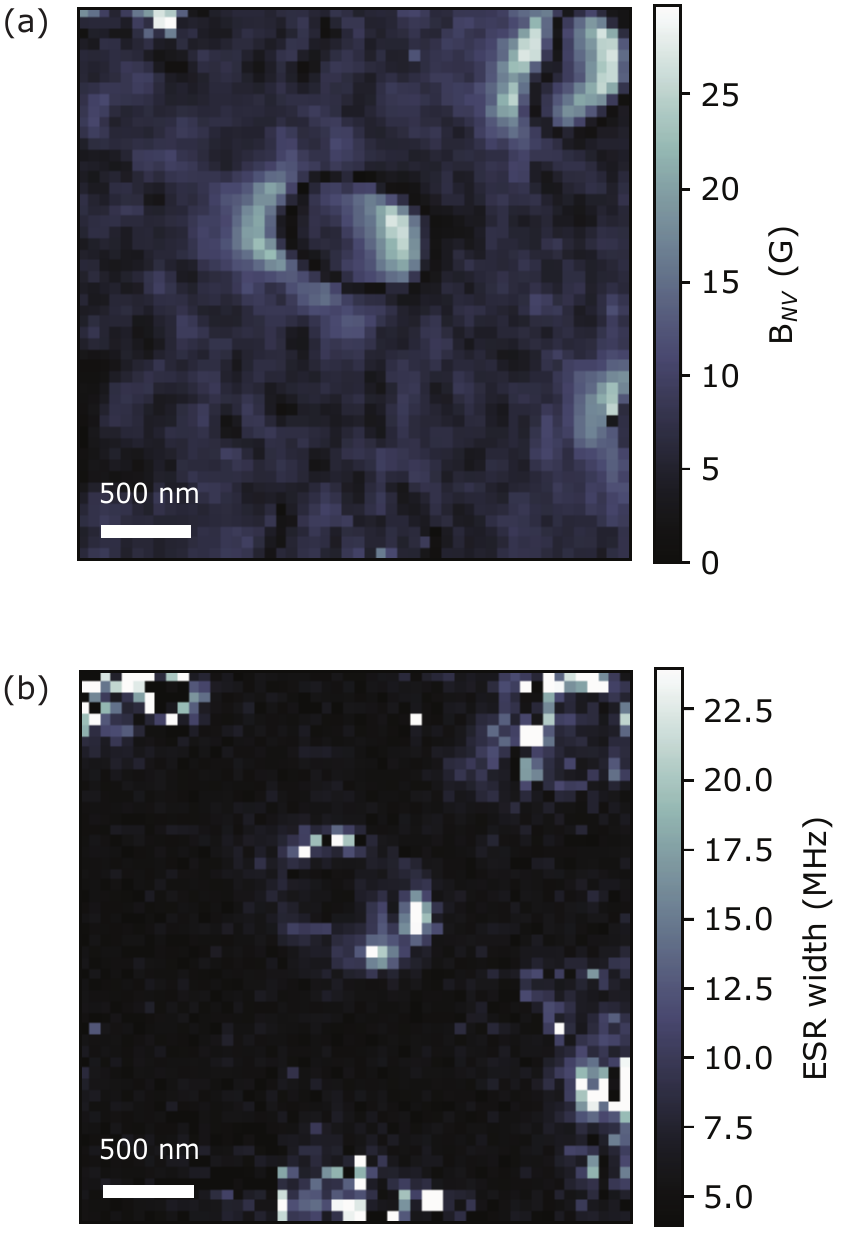}
\caption{\label{fig4} Enhanced magnetic fluctuations near skyrmion domain walls (a) Spatial map of the magnetic field along the NV axis, $B_{NV}$, showing isolated skyrmions. Dark contours indicate the location of the domain wall. The external field, $B_z = 9.5$ Oe, is normal to the film plane. (b) NV ESR width averaged over both ESR dips, showing enhanced mangetic fluctuations at the skyrmion domain wall.
}
\end{figure}

Importantly, our NV imaging technique allows us to glean quantitative information about the time scale of the domain wall dynamics. We estimate the average hopping rate, $\chi$, of the domain wall in Fig 3h to lie between 60 Hz and 14 MHz by making the following two observations. First, telegraph switching of the NV fluorescence rate was not observed on time scales slower than 16 ms, putting 60 Hz as a lower bound on $\chi$. In this experiment, we could not explore faster time scales because of insufficient signal to noise ratio for measurement time bins shorter than 8 ms. An upper limit on the characteristic hopping frequency can be set by treating the domain wall position between the two pinning sites as a quasi-1D system and assuming that the dynamics is governed by an Arrhenius type thermal activation of hopping between two sites \cite{Eltschka2010}, where the number of domain wall jumps in a given time is described by a Poisson process. In this case, the NV ESR spectrum is expected to take two different forms, displaying either a single resonance line or a split pair of resonance lines for each spin transition, depending on the characteristic rate of domain wall hopping $\chi$ and the corresponding spectral shift of the NV ESR dip. Focusing on one NV spin transition, for example $m_s=0\rightarrow-1$, the spectral shift can be written as $\Delta f_{ESR} = f_{DW2} - f_{DW1}$, where $f_{DW1}$ and $f_{DW2}$ are the NV resonance frequencies corresponding to the two domain wall positions. In the limit $\Delta f_{ESR}/2 > \chi$, distinct resonances will be observed at $f_{DW1}$ and $f_{DW2}$, whereas in the limit $\Delta f_{ESR}/2 < \chi$, an effect similar to motional narrowing will give a spectrum with a single resonance at $(f_{DW2} + f_{DW1}$)/2 if, on-average, an equal amount of time is spent in both domain wall positions \cite{Bloembergen1948, Li2013}. The four distinct ESR lines shown in Fig. 3b indicate that the first limit applies and that the characteristic frequency of the observed domain wall hopping is limited to $\chi<14$ MHz. This reasoning can also be applied to other observed bistable skyrmion bubble walls, for example those shown in Fig. 3c-e. Assuming that these domain wall branchings are due to a similar hopping mechanism observed in Fig. 3h-m, the fact that two domain wall positions are observed in Fig. 3c-e can be used to place a rough limit on the timescale of those domain wall dynamics as well. This novel functionality of the NV center presents an opportunity in the future to probe the dynamics more finely. Repeating spectral measurements at a different distances from the hopping bubble walls would allow one to probe frequency scales down to the NV ESR width.

In addition to large jumps of domain wall position that produce a discrete set of ESR splittings, we also observe smaller magnetic fluctuations that broaden the NV spin transitions when the NV is positioned near a domain wall. Fig. 4b shows a spatial map of the average ESR width and comparison with a stray field image taken in the same area (Fig. 4a) clearly shows that magnetic fluctuations are enhanced near all skyrmion bubble walls, even in the absence of the clear bistabilities seen in Fig 3. We note that this broadening is not due to fluctuations in the NV-domain wall distance in regions of high magnetic fields gradient, as it is not observed near other sharp magnetic features not associated with bubble domain walls (supplementary section S3). The emergence of these enhanced fluctuations near domain walls is not well understood and can be interpreted in a few ways. It could imply the existence of small fluctuations in the position of all domain walls, driven thermally or magnetically \cite{Kronseder2015}, possibly by the applied microwaves or laser light \cite{Tetienne2014a}. Alternatively, the spatial dependence seen in Fig. 4b could be the result of amplification or concentration of magnetic noise sources, such as spin waves, near the domain walls \cite{Mochizuki2014, Wagner2016}.

\section{\label{sec:structure}Skyrmion bubble structure}

\begin{figure*}
\includegraphics[width=\textwidth]{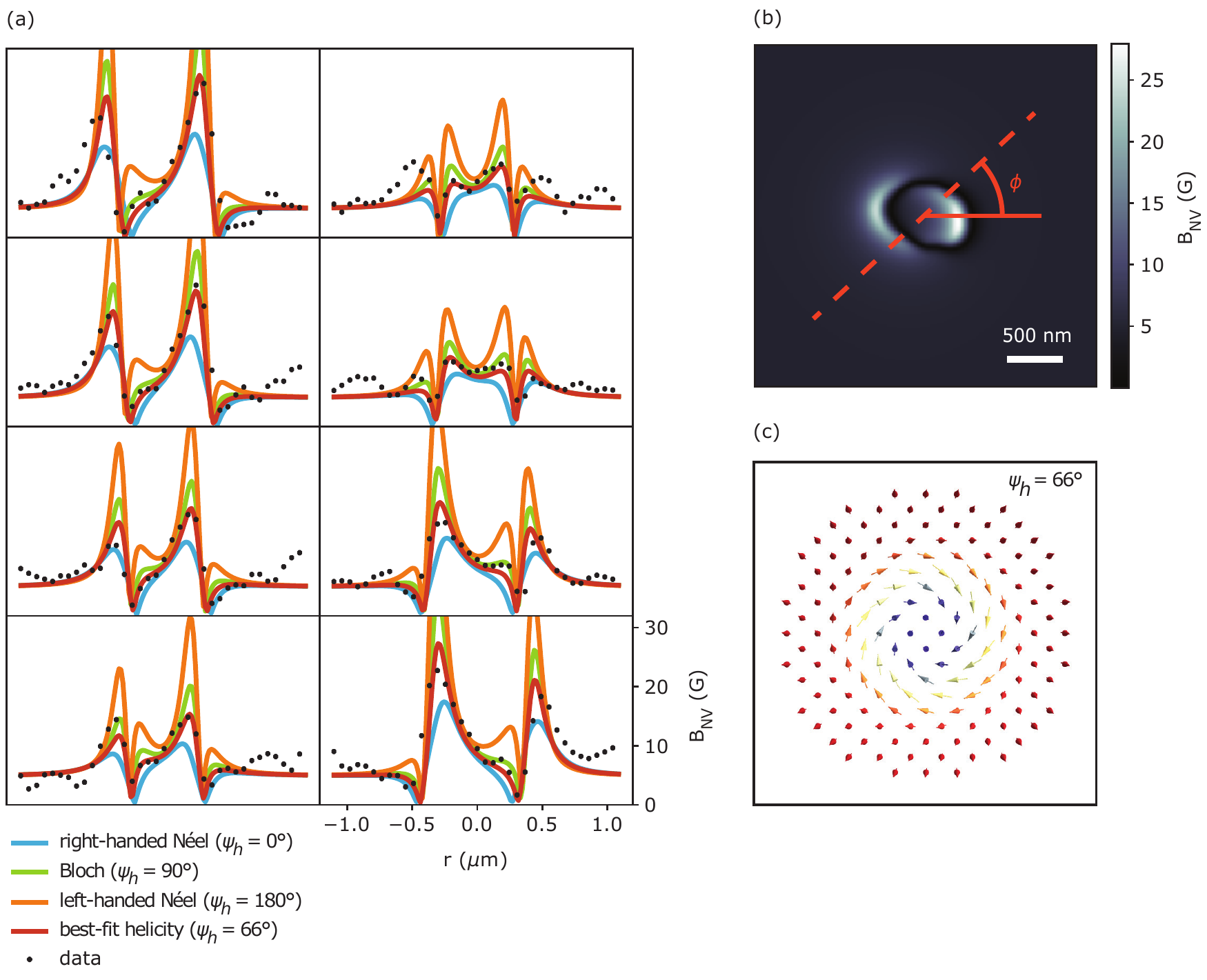}
\caption{\label{fig5} Reconstruction of helicity angle. (a) Linecuts at various angles $\phi$ across the magnetic bubble, comparing the measured magnetic field shown in Fig. 4a to the simulated field for four domain wall types--- right-handed N\'eel, Bloch, left-handed N\'eel, and a domain wall with $\psi_h=66^{\circ}$. (b) The simulated magnetic field along the NV axis for the best fit helicity angle $66^{\circ}$. (c) Schematic of the magnetization of a skyrmion bubble with a fixed helicity angle $\psi_h=66^\circ$, viewed from above. The magnetization transitions from pointing upward outside the bubble (red) to pointing downward inside (blue) with a rotation direction between that of Bloch and N\'eel-type domain walls.
}
\end{figure*}

The magnetic structure of skyrmions has important implications for the viability and design of skyrmion-based devices because the structure determines important parameters of current-driven skyrmion motion, such as the skyrmion Hall angle and velocity \cite{Jiang, Neubauer2009, Tomasello2014, Litzius2017}. However, probing the structure of skyrmion bubble is difficult due to the required nanometer-scale resolution. While many techniques can be used to determine domain wall structure in principle, there are few examples of probes that are local, non-invasive, and capable of studying a wide range of skyrmion materials. Recent NV imaging studies of magnetic thin films have established scanning NV microscopy as a useful probe of magnetic structure with all these features \cite{Dovzhenko2018, Gross2016, Tetienne2015, Hingant2015}. Starting from a map of the NV-axis stray field (Fig. 4a), a quantitative reconstruction of the magnetization structure is possible but requires knowledge of several material parameters, careful calibration of the NV scan height, and some structure assumptions based on micromagnetic theory. In this work, we determine all relevant materials parameters, leaving free only the domain wall helicity angle $\psi_h$. The helicity angle sets the rotation direction of magnetization through a cross-section of the domain wall--- here defined relative to the common domain wall types as $\psi_h = 0, \pi/2$, and $\pi$ for right-handed N\'eel, Bloch, and left-handed N\'eel respectively. With this approach we can use the local, nanoscale nature of the NV probe to search for variations in the helicity angle along the skyrmion domain wall, which allows us to check for a fixed chirality of individual bubbles.

We start by assigning a polarity to regions of the $B_{NV}$ map separated by zero-field contours. The polarity direction is determined by the direction of the applied external field. This signed field map can in turn be used to calculate the full vector components of the stray magnetic field \cite{Dovzhenko2018,Lima2009}. The $z$ component of $\mathbf{B}$ (where $\mathbf{\hat{z}}$ is normal to the sample) at the sample surface can be extrapolated and used to estimate the domain wall position (supplementary section S4). The magnetization pattern $\mathbf{M}$ is then fully determined by the domain wall width $\Delta_{DW}$, saturation magnetization $M_s$, and helicity angle $\psi_h$.

For NV-sample separations larger than $\Delta_{DW}$, a direct measurement of $\Delta_{DW}$ is difficult and $\Delta_{DW}$ must be inferred from measurements of other parameters. Three parameters are required--- $M_s$, PMA energy density $K_{eff}$, and domain wall energy density $\gamma_{DW}$. The domain wall width is given by $\Delta_{DW} =\sqrt{A_{ex}/K_{eff}}$ \cite{Rohart2013,Meier2017}, corresponding to a magnetization profile across the domain wall $M_z=M_s \text{tanh}(x/\Delta_{DW})$. First $M_s$ and $K_{eff}$ are measured with a SQUID magnetometer, then $\gamma_{DW}$ is directly measured with scanning NV images of the stripe phase. The domain wall energy density can be estimated from the period of the stripe spacing in these images, or calculated more directly by comparing the demagnetization energy and total domain wall length in a an image area. The exchange stiffness is then calculated from $\gamma_{DW}=4\sqrt{A_{ex} K_{eff}}-\pi |D|$, where $D$ is the DMI energy density determined from Brillouin light scattering (BLS) measurements (supplementary section S5) \cite{Ma2016}. Armed with these parameters, we can compare the expected stray field for a given helicity to that measured with the scanning NV center.

Figure 5b shows the best-fit simulated stray field corresponding to the measurement stray field in Fig. 4a. The helicity angle, $\psi_h = 66^{\circ}$ is determined by a 2-dimensional fit to simulated data in 1.4 $\mu$m box around the stray field features of the central skyrmion. Values obtained for other skyrmions include $\psi_h=73^{\circ}$ and $75^{\circ}$ (supplementary section S6). To allow for the possibility that the helicity angle can change locally along the bubble domain wall, it is instructive to compare the simulated stray field as a function of position along the domain wall. In Fig. 5a, linecuts across the measured and simulated field images are shown as a function of cut angle and helicity type. The skyrmion bubbles in this material consistently show a slightly right-handed helicity angle. A constant right-handed helicity is consistent with a non-zero winding number, but it is important to note that uncertainties in sample thickness and NV height will change the measured helicity angle. The right-handed helicity observed here agrees with BLS measurements of $D$, but micromagnetic theory gives a smaller helicity angle for the measured $D$ (supplementary section S4).

\section{\label{sec:discussion}Discussion and future NV studies of skyrmions}

This work extends recently developed scanning NV microscopy techniques used to study multilayer skyrmion materials. Specifically, we have demonstrated that NV microscopy can simultaneously locally probe both magnetic structure and dynamics. As shown in similar thin film systems \cite{Gross2018}, pinning and disorder are important factors in determining static skyrmion bubble sizes in Ta/CoFeB/MgO. We have shown that these sizes are determined dynamically, via hopping of skymion bubble domain walls between pinning sites. We have also probed the dynamics of these hopping processes and we've observed ubiquitous magnetic fluctuations near skyrmion bubble domain walls. Our measurements confirm the right-handed helicity of the DM interaction in this specific material structure and we have measured helicity angles in a range $\psi_h = 66^{\circ}-75^{\circ}$.

Our images highlight the importance of pinning interactions between defects and the internal degrees of freedom of skyrmion bubbles. As shown previously \cite{Woo2016}, standard defect-agnostic micromagnetic models of skyrmion motion break down when material inhomogeneities exist on length scales smaller than the skyrmion size. In this work, we experimentally observe the interaction of skyrmion bubbles with these inhomogeneities. This indicates that the behavior of future devices based on the skyrmion bubbles in this material, or similar materials, will be determined by pinning interactions. Specfically, these pinning sites will likely determine the current-driven velocity and trajectory of skyrmion motion \cite{Woo2016,Litzius2017}. For use in future devices, skyrmions with smaller diameters are desired \cite{Fert2017} and as the development of multilayer skyrmion material continues, the high spatial resolution of scanning NV microscopy will be increasingly important for the characterization of nanometer-scale skyrmions.

Inspired by this work, a more thorough study of the fluctuation dynamics observed here is a promising direction for future NV-based skyrmion experiments. NV noise spectroscopy has been developed as a powerful tool for obtaining information about the dynamics of noise processes in materials \cite{Agarwal2017,Myers2014a,Myers2017a,DeLange2010,VanderSar2015,Du2017a}. In the future, NV noise spectroscopy could be utilized to study thermal fluctuation dynamics, which are thought to play an important role in the current driven motion of skrymions \cite{Troncoso2014}.

\begin{acknowledgments}
We thank Preeti Ovartchaiyapong for diamond fabrication advice and Simon Meynell and Susanne Baumann for helpful discussions. The work at UCSB was supported by an Air Force Office of Scientific Research PECASE award. A portion of the work was done in the UC Santa Barbara nanofabrication facility, part of the NSF funded NNIN network. The authors acknowledge support from the Nanostructures Cleanroom Facility (NCF) at the California NanoSystems Institute (CNSI). Use of the Shared Experimental Facilities of the Materials Research Science and Engineering Center at UCSB (MRSEC NSF DMR 1720256) is gratefully acknowledged. The UCSB MRSEC is a member of the NSF-supported Materials Research Facilities Network (www.mrfn.org). Kang L.
Wang acknowledges the support of NSF-1611570. Guoqiang Yu acknowledges the financial support from the National Natural Science Foundation of China [NSFC, Grants No.11874409], the National Natural Science Foundation of China (NSFC)-Science Foundation Ireland (SFI) Partnership Programme [Grant No. 51861135104], and 1000 Youth Talents Program. Xin Ma acknowledges the support from NSF grant CCF 1740352 and SRC nCORE NC-2766-A. The work at UT Austin was primarily supported as part of SHINES, an Energy Frontier Research Center funded by the U.S. Department of Energy (DOE), Office of Science, Basic Energy Science (BES) under Award No. DE-SC0012670.
\end{acknowledgments}

\bibliography{tacofeb_references_f} 

\clearpage
\setlength{\jot}{18pt}
\renewcommand{\figurename}{Fig. S}
\setcounter{figure}{0}

\onecolumngrid

\begin{center}
{\Large Supplementary Information: Single spin sensing of domain wall structure and dynamics in a thin film skyrmion host}
\end{center}

\section*{\label{sec:S1}S1. Contour imaging— domain wall position}

\begin{figure}[h]
\includegraphics[width=\textwidth]{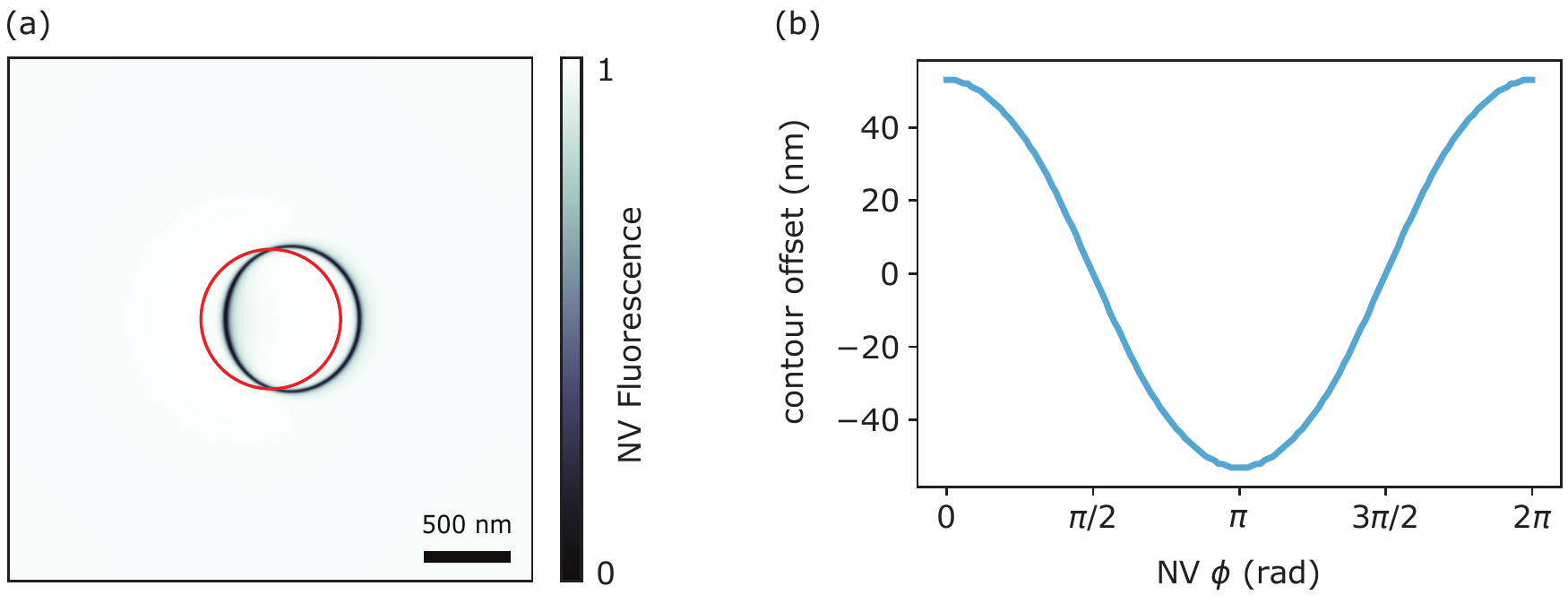}
\let\nobreakspace\relax
\caption{ (a) Simulated zero-field contour image from a circular domain wall in Ta/CoFeB/MgO with helicity angle $\psi_h=62^{\circ}$. The red circle corresponds to the center of the domain wall and the black circle is the simulated NV contour signal. The external field is 9.5 Oe. (b) Simulated offset between 2870 MHz contour and domain wall position as a function of the in-plane NV angle relative to the domain wall normal. In both plots the NV angle and height are the same as those found for NV used to produce Fig. 4 of the main text.
}
\label{contourShift}
\end{figure}

The NV contour imaging method consists of fixing an applied microwave frequency while monitoring the NV fluorescence or the ratio of NV fluorescence for microwaves on vs.\ off. Dark contours in a contour scanning NV image correspond to locations where the stray field from the sample cause NV spin transitions to line-up with the applied microwave frequency. As stated in the main text, at low external fields and with magnetic domain wall widths much smaller than the bubble or stripe domains, 2.870 GHz contour images give good approximations to the domain wall positions. Figure S1a shows a simulated zero-field contour from a circular domain wall with a 9.5 Oe external magnetic field. The zero-field contour is shifted slightly along the direction of the NV center's in-plane projection. Figure S\ref{contourShift}b shows the offset (in nm) between the position of a straight Bloch domain wall and its zero-field contour in the imaging plane as a function of the in-plane NV angle relative to a direction normal to the domain wall ($B_{ext}=0$).

\section*{\label{sec:S2}S2. NV height calibration}

The NV scan height above the magnetic CoFeB layer determines the spatial resolution of the NV imaging technique and this height an important parameter in reconstructing the magnetization. The NV height is calibrated by scanning the NV across a step edge that has been etched through the magnetic layer \cite{Tetienne2015, Hingant2015}. The step edge is defined with electron beam lithography and the thin-film stack is etched with Ar ion milling. An external magnetic field is applied to saturate the film and the stray field is measured as a function of position from the step edge. Assuming a sharp step edge, with a magnetic layer thickness $t$ much less than the NV height $h$, the stray field profile is given by 
\begin{align}
\label{eq1}
B_x &= \frac{\mu_0 M_s t}{2\pi}\frac{h}{(x-x_0)^2+h^2} \\
\label{eq2}
B_z &= -\frac{\mu_0 M_s t}{2\pi}\frac{x}{(x-x_0)^2+h^2} + B_{z,0}
\end{align}
where $B_{z,0}$ is the external magnetic field and the step edge runs along the line $x=x_0$.

\begin{figure}[t]
\includegraphics[width=0.7\columnwidth]{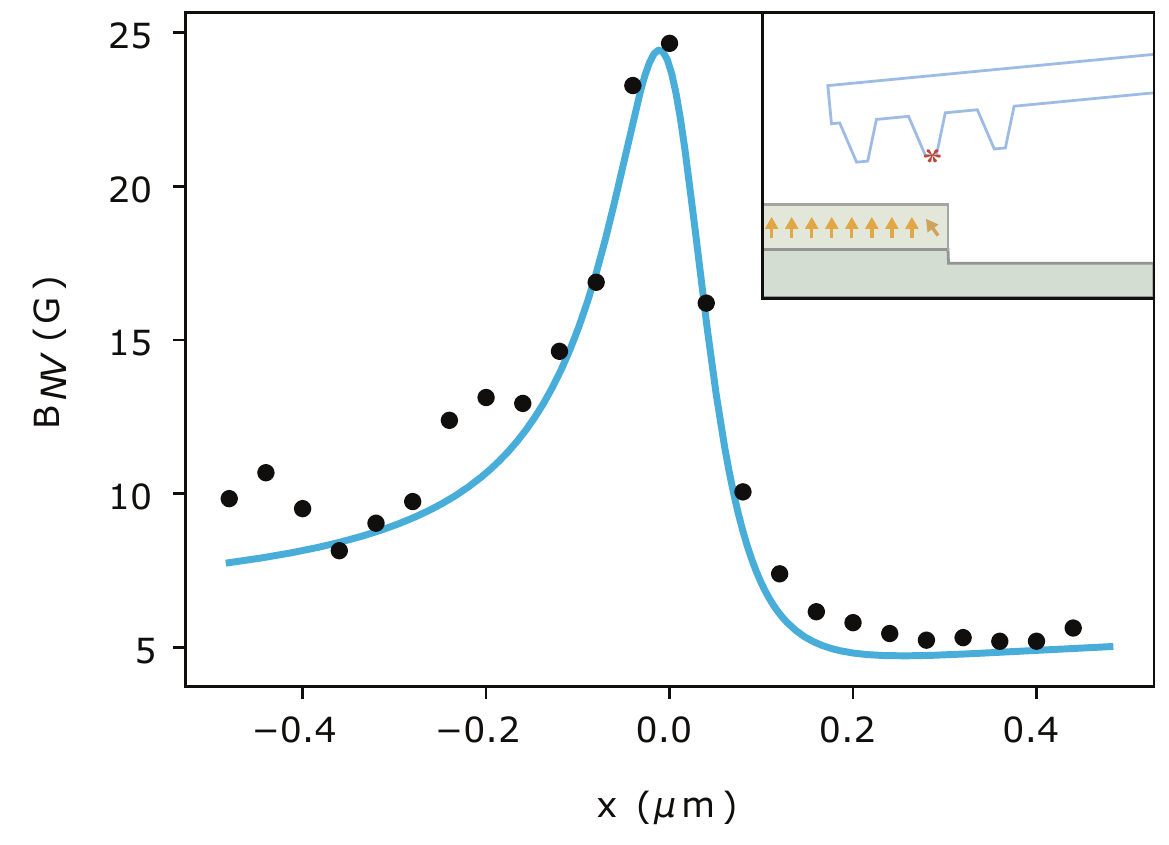}
\let\nobreakspace\relax
\caption{NV height calibration curve example. $B_{NV}$ is measured as a function of position across a domain wall (black). Only two parameters, NV height and lateral domain wall position, are left free in the fit (blue). (inset) NV height calibration diagram. NV scan height is determined by tip-sample tilt and location of the NV in a small array of pillars (typically 3x3 or 7x7 pillars, with 2 $\mu$m separation).}
\label{calibrationCut}
\end{figure}

In the presence of non-zero DMI, the magnetization rotates at the step edge \cite{Tetienne2015}. This rotation angle at the edge is given by \cite{Rohart2013}
\begin{align}
\theta_0 = \text{arcsin}\left(\frac{2 \Delta_{DW} D}{A_{ex}}\right)
\end{align}
and is described by
\begin{align}
\left(\frac{d\theta}{dx} \right)^2 = \frac{C+\text{sin}^2(\theta)}{\Delta^2_{DW}}
\end{align}
which for small values of $D$ gives,
\begin{align}
\theta &\approx \theta_0 e^{-|x|/\Delta_{DW}} \\
M_x &\approx M_s \frac{2 \Delta_{DW}D}{A_{ex}}e^{-|x|/\Delta_{DW}} \\
M_z &= \sqrt{M^2_s - M^2_x}
\end{align}

Taking this rotation into account, the stray field can be calculated as
\begin{align}
B_x &= -\frac{\mu_0 t}{2\pi}\int_{-\infty}^0 dx_0 \left\{ \frac{M_x}{(x-x_0)^2 + h^2} - \frac{2(x-x_0)(M_x(x-x_0)+M_z h)}{((x-x_0)^2 + h^2)^2} \right\} \\
B_z &= -\frac{\mu_0 t}{2\pi}\int_{-\infty}^0 dx_0 \left\{ \frac{M_x}{(x-x_0)^2 + h^2} - \frac{2h(M_x(x-x_0)+M_z h)}{((x-x_0)^2 + h^2)^2} \right\}
\end{align}

For the low DMI Ta/CoFeB structure studied here, this magnetization rotation at the step edge gives only a few nm correction to the NV height calibration. The NV height above the CoFeB layer is extracted by fitting the stray field given by these expressions to the measured field along the NV axis. The magnetic moment density $M_s t$ is measured separately with SQUID (see Sec. S2). An example of one of the calibration curves is shown in Fig. S\ref{calibrationCut}. Twenty of these linecuts were taken at different points along a 2 $\mu$m length of the step edge. The NV height and height error used in magnetization reconstruction calculations is given by the mean and standard deviation of these fits respectively.

The angles of the NV axis with respect to the edge and sample normal are calibrated separately using an external magnetic field. The NV height calibration depends critically on the profile of the magnetization at the etched edge. Any redeposition of magnetic material or damage to the magnetic structure at the edge induced by ion milling will lead to systematic errors in the NV height.

\section*{\label{sec:S3}S3. ESR widths at etched step edge}

\begin{figure}[h]
\includegraphics[width=0.7\columnwidth]{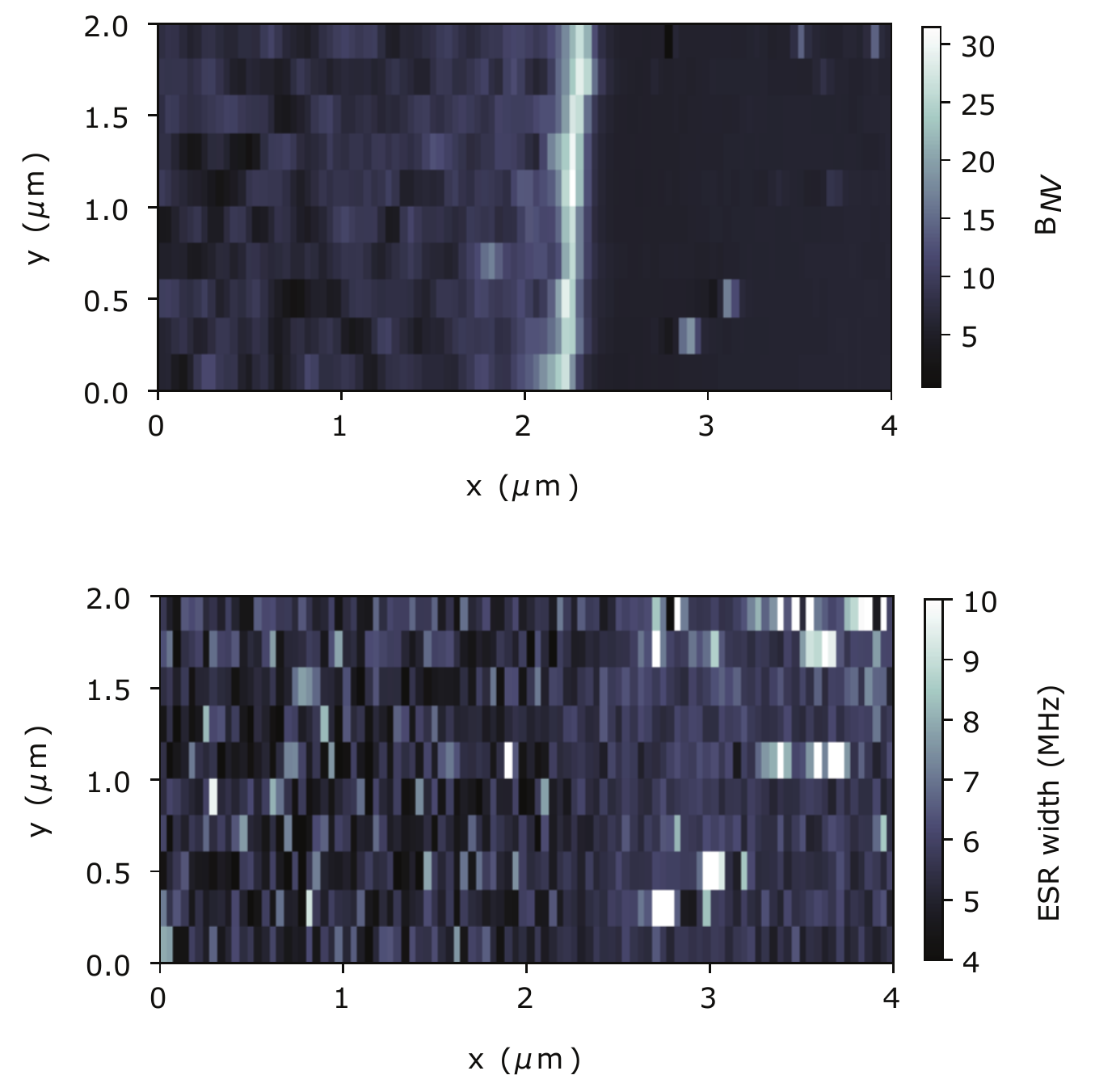}
\let\nobreakspace\relax
\caption{Stray magnetic field along the NV axis measured from the ESR spectrum (top) and average ESR resonance width of each spectrum (bottom). The large magnetic field in the center of the top image corresponds to the step edge position. There is no increase in ESR width at the step edge relative to the saturated magnetic state (left side). The isolated magnetic features and ESR broadening in the etched region (right side) are due to magnetic islands that were masked during the step edge etch.}
\label{edgeESR}
\end{figure}

To verify that the ESR broadening observed near domain walls (Fig. 4 from the main text) results from magnetic noise intrinsic to the domain wall and is not, for example, due to tip oscillation in a high field gradient, the NV ESR width is measured as the NV is scanned across a step edge etched in the magnetic thin film. Figure S3 plots the stray magnetic field and the ESR width in the same imaging area. The large gradient above the step edge (Fig. S\ref{edgeESR} top, bright vertical line) does not have an associated increase in ESR width (Fig. S\ref{edgeESR} bottom), thus confirming that NV motion in a high field gradient is not responsible for the ESR broadening observed near skyrmion domain walls in Fig 4 of the main text.

\section*{\label{sec:S4}S4. Extracting material parameters}

A quantitative comparison of a skrymion's stray field to its underlying magnetic structure necessitates knowledge of several material parameters, as pointed out in the discussion of Fig. 5 in the main text. Specifically we require a knowledge of $M_s$, $t$, and domain wall width $\Delta_{DW}$, and we assume an analytic form (see section S5) for the domain wall profile. The domain wall position is also needed and is given by the NV images. To estimate $\Delta_{DW}$, which cannot be measured directly with our NV center for lack of spatial resolution, we combine bulk measurements of the effective magnetic anisotropy energy density $K_{eff}$ and the magnetic surface density $I_s = M_s t$ with NV measurements of the domain wall energy density $\gamma_w$. 

\begin{figure}[h]
\includegraphics[width=0.7\columnwidth]{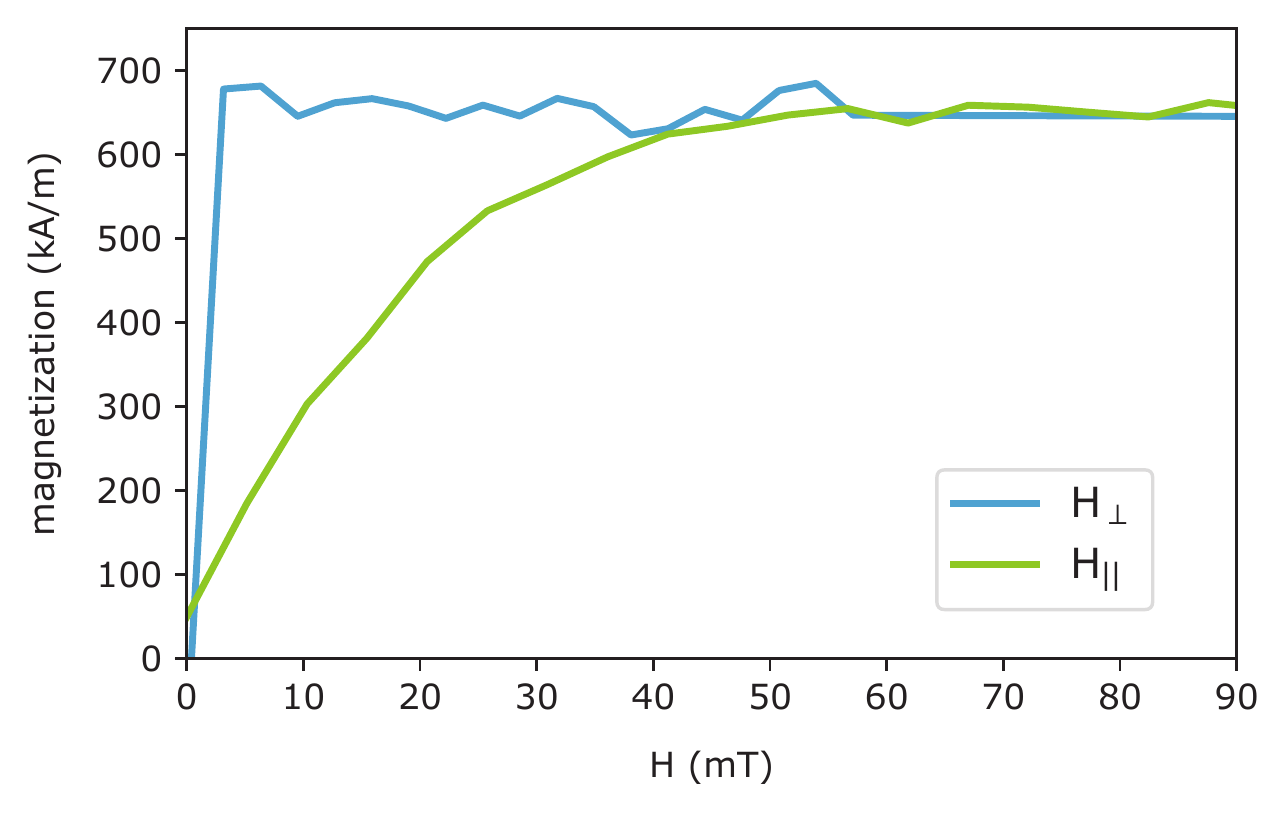}
\let\nobreakspace\relax
\caption{SQUID measurement of the saturation magnetization, $M_s$, and perpendicular magnetic anisotropy, $K_{eff}$. The sample magnetization is measured as function of external magnetic field, applied normal to the film plane (blue) and parallel to the film plane (green). The total sample moment is measured, and the magnetization is calculated assuming a CoFeB thickness of 1 nm.}
\label{SQUID}
\end{figure}

The parameters $K_{eff}$ and $M_s$ are calculated from hysteresis curves obtained with a Quantum Design MPMS 5XL SQUID. The magnetic moment is measured as a function of an applied field both parallel and normal to the film plane (Fig. S\ref{SQUID}). After dividing the measurement magnetic moment by the thin film volume, $M_s$ is the saturation value, while $K_{eff}$ is given by the area between the parallel and normal curves \cite{Johnson1996}. These measurements give $I_s = (6.5 \pm 0.1) \times 10^{-4}$ A, where the uncertainty is given the standard deviation of the saturated SQUID value. This gives $M_s = 6.6 \times 10^5$ A/m and $K_{eff} = 8.3 \times 10^3$ J/m$^3$ for a film thickness $t=1.0$ nm. Calculating the uncertainty in $M_s$ and $K_{eff}$ is trickier because their values depend on the film thickness $t$. The value of $t$ used in these calculations is given by the measured deposition thickness, but we note that magnetic dead layers have been observed in CoFeB films at Ta or MgO interfaces \cite{Jang2010,Sinha2013}. A non-zero dead layer thickness would lead to different values of $M_s$ and $K_{eff}$ and differences in the magnetization reconstruction.
 
\begin{figure}
\includegraphics[width=\columnwidth]{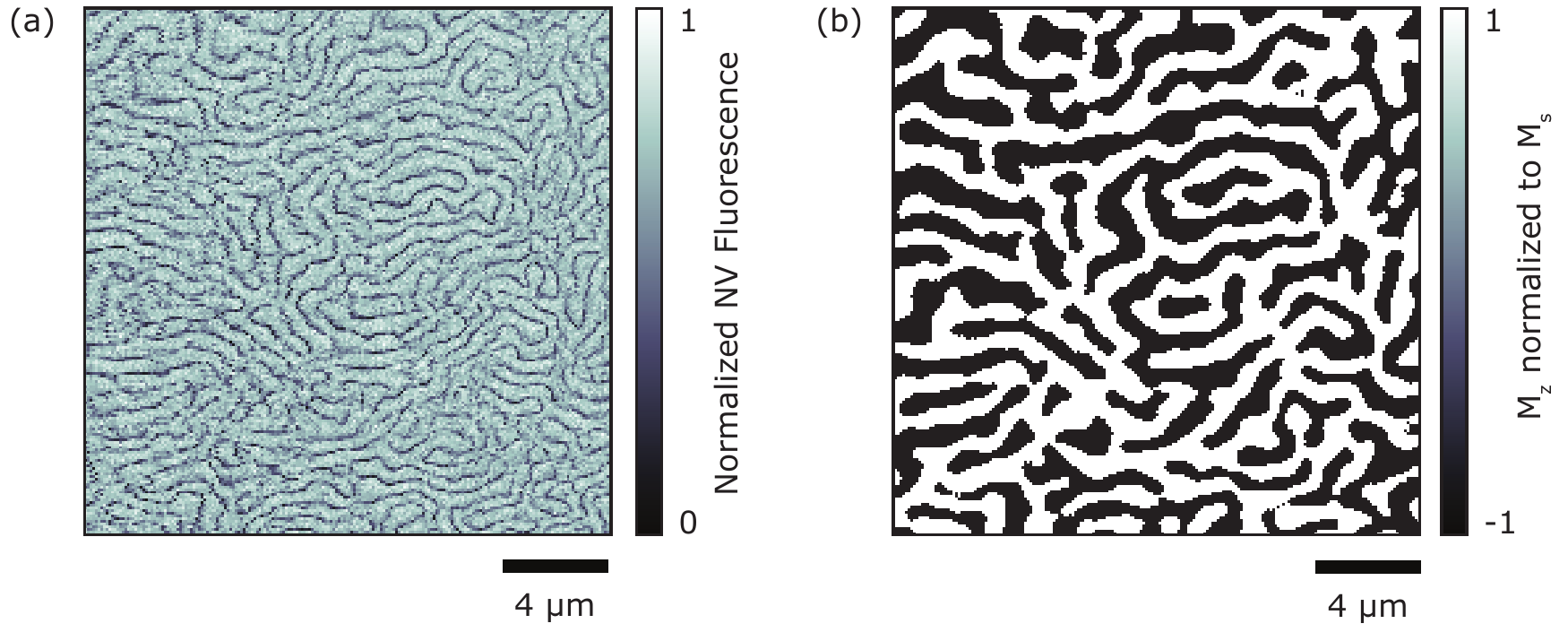}
\let\nobreakspace\relax
\caption{Stripe domains at zero external magnetic field. (a) Measured zero-field contours and (b) corresponding assigned magnetization used to calculate the demagnetization energy.}
\label{stripeDomains}
\end{figure}

Armed with values of $M_s$ and $K_{eff}$, it is possible to measure the domain wall energy density $\gamma_w$ from stripe-phase NV images. The NV scan can be used to calculate the demagnetization energy and total domain wall length in the image area. The domain wall energy density is then calculated based on an energy minimization that balances variation in the demagnetization energy and domain wall length measured for a particular imaging area \cite{Buford2016}. Starting from a zero-field stripe image, a polarity ($\pm M_s$) is assigned to the two image regions (Fig. S\ref{stripeDomains}). The demagnetization energy $E_{demag}$ in the imaged area is calculated with OOMMF \cite{Donahue1999} using mirror symmetric boundaries. The total length of the domain walls in this image $L$ is also measured and both values are normalized by the image area giving $E'_{demag}$ and $L'$. The domain wall energy density can then be obtained as in \cite{Buford2016} by calculating the variation of $E'_{demag}$ and $L'$ with respect to the characteristic stripe periodicity $p$,
\begin{align}
\gamma_w = -\frac{\partial E'_{demag}/\partial p}{t\partial L'/\partial p}
\end{align}
These variations can be related to variations in the imaging resolution,
\begin{align}
\frac{\partial E'_{demag}}{\partial p} = \frac{ E'_{demag}(s+\Delta s) - E'_{demag}(s)}{\Delta s}
\end{align}
\begin{align}
\frac{\partial L'}{\partial p} = \frac{ L'}{a}
\end{align}

This procedure gives $\gamma_w = 1.3$ mJ/m$^2$. The domain wall energy density can be used in turn to calculate $A_{ex}$ and $\Delta_{DW}$
\begin{align}
A_{ex} = \frac{(\gamma_w + \pi D)^2}{16 K_{eff}} \\
\Delta_{DW} = \sqrt{A_{ex}/K_{eff}}
\end{align}

The DMI strength measured via Brillioun light scattering (BLS) is found to be $D=47$ $\mu$J/m$^2$, giving an exchange stiffness $A_{ex}=9$ pJ/m and domain wall width $\Delta_{DW}=33$ nm. The extracted value of $A_{ex}$ is in good agreement with the value measured with BLS (section S5). This method of estimating $\gamma_w$ can be checked against an analytic form for parallel stripe domains \cite{Saratz2010}
\begin{align}
\gamma_w = \frac{\mu_0 M_s^2 t}{\pi}\left( \text{ln}\left( \frac{2L}{\pi t} \right) + \frac{1}{2} \right)
\end{align}

Identifying characteristic image length scale in the image Fourier decomposition, $L=664 \pm 23$ nm. This gives a domain wall width $\Delta_{DW} = 35$ nm.

However, a non-zero dead layer thickness will alter either calibration of domain wall width. For example, assuming a dead layer thickness of 0.36 nm, similar to \cite{Jang2010}, the same analysis gives the parameters $M_s = 9.5 \times 10^5$ A/m, $K_{eff} = 1.2 \times 10^4$ J/m$^3$, $A_{ex} = 20$ pJ/m, and $\Delta_{DW} = 49$ nm.

\section*{\label{sec:S5}S5. BLS measurements of material parameters}
\subsection*{Spin wave dispersion in BLS measurements:}
The spin waves probed here are Damon-Eshback (DE) modes with propagation directions perpendicular to external magnetic field. The spin wave dispersion is described by \cite{Ma2017}:
\begin{align}
\begin{split}
f = &\frac{\gamma}{2\pi}\sqrt{\left( H+\frac{2A_{ex}}{M_s}k^2 + 4\pi M_s (1-\xi(kL)) - \frac{2k_\perp}{M_s} \right)\left(H+\frac{2A_{ex}}{M_s}k^2 + 4\pi M_s \xi (kL)\right)}\\
&+\epsilon(\mathbf{H}_{EB},K_\perp , k)*sgn(k M_z) - sgn(M_z)\frac{\gamma}{\pi M_s} D k
\label{blsEQ1}
\end{split}
\end{align}
where $H$ is the magnitude of the external magnetic field, $\gamma$ is the gyromagnetic ratio, $A_{ex}$ is the exchange stiffness constant, $\xi(kL) = 1-(1-e^{-|kL|})/|kL|$ with $L$ the CoFeB thickness, $K_\perp$ is the interfacial magnetic anisotropy which mainly originates from the CoFeB/MgO interface, $\epsilon(\mathbf{H}_{EB},K_\perp , k)$ describes a correction in frequency due to the non-reciprocity of the DE spin waves, and $D$ is the DMI coefficient.

\begin{figure}[h]
\includegraphics[width=\columnwidth]{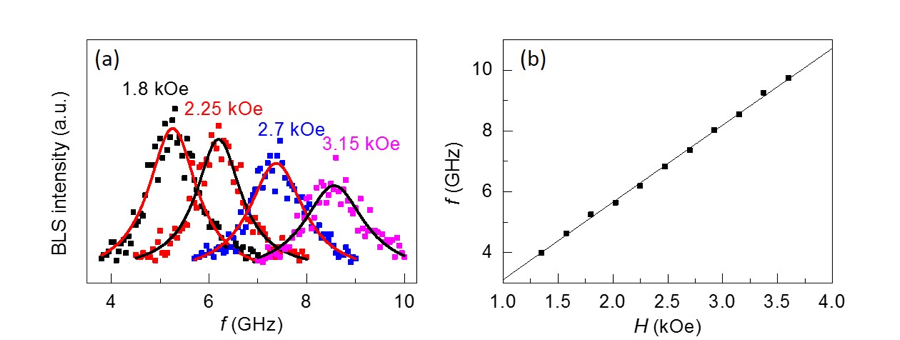}
\let\nobreakspace\relax
\caption{(a) BLS spectra with different $H$ field applied and (b) the dependence of $f$ on $H$.}
\label{BLS1}
\end{figure}

\subsection*{Field-dependent BLS measurements to determine magnetic anisotropy:} 
Magnetic field dependent BLS measurements were performed to determine the magnetic anisotropies. In order to improve the accuracy of magnetic anisotropy measurements, we performed BLS measurements with normal incidence of light ($k\approx 0$). As a result, Eq. \ref{blsEQ1} is simplified to $f=\frac{\gamma}{2\pi}\sqrt{H(H+4\pi M_{eff})}$ with $4\pi M_s - \frac{2K_\perp}{M_s}$. Figure S\ref{BLS1}(a) shows that the BLS spectra under different $H$ for the Ta/CoFeB/Pt/MgO sample. The BLS spectra can be well fitted with Lorentzian functions, and the resonance frequency $f$ increases with $H$. Figure S\ref{BLS1}(b) displays $f$ as a function of $H$, which can be well fitted by the simplified Eq. \ref{blsEQ1} at $k=0$ with $4\pi M_{eff}=-0.27$ kOe (solid lines). The negative sign of $4\pi M_{eff}$ indicates that the easy axis of magnetization is perpendicular to the thin film plane. 

\begin{figure}[h]
\includegraphics[width=\columnwidth]{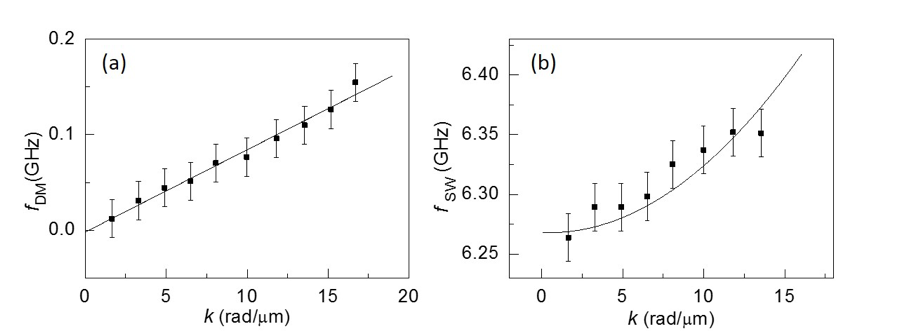}
\let\nobreakspace\relax
\caption{The dependence of $f_{DM}$ and $f_{SW}$ on $k$.}
\label{BLS2}
\end{figure}

\subsection*{k-dependent BLS measurements to determine DMI coefficient $D$ and exchange stiffness $A_{ex}$:}
Both the DMI coefficient $D$ and exchange stiffness constant $A_{ex}$ can also be determined from the momentum-resolved BLS measurements. 
We determine the $D$ by subtracting the $f(k)$ and $f(-k)$ in Eq. \ref{blsEQ1}:
\begin{align}
f_{DM} = \frac{1}{2} \left( (f(-k,M_z)-f(k,M_z)) - (f(-k,-M_z)-f(k,-M_z)) \right) = \frac{2\gamma}{\pi M_s} D k
\label{blsEQ2}
\end{align}  
Figure S\ref{BLS2} shows that the linear correlation between $f_{DW}$ and $k$, where the slope is used to determine $D=47~ \mu$J/m$^2$. The positive sign of $D$ indicates that the right-handed magnetic chirality is preferred in the material system.
The exchange stiffness $A_{ex}$ is determined by averaging $f(k)$ and $f(-k)$ in Eq. 
\ref{blsEQ1}:
\begin{align}
\begin{split}
f_{SW} =& \frac{1}{2} \left( f(k)+f(-k) \right)\\ =& \frac{\gamma}{2\pi} \sqrt{\left(H+\frac{2A_{ex}}{M_s}k^2 + 4\pi M_s (1-\xi(kL)) - \frac{2 k_\perp}{M_s} \right) \left( H+\frac{2A_{ex}}{M_s}k^2 + 4\pi M_s \xi(kL) \right)}
\label{blsEQ3}
\end{split}
\end{align}  
Figure S\ref{BLS2}(b) plots the $f_{SW}$ as a function of $k$, where $f_{SW}$ increases with larger $k$ and can be well fitted with the above simplified equation. As a result, we derived the exchange stiffness $A_{ex}=7.8$ pJ/m.

\section*{\label{sec:S6}S6. Reconstructing helicity angle}

Constructing the underlying magnetization pattern from a stray field measurement is not trivial. The mapping from magnetization to stray field is not one-to-one and cannot be inverted. In order to say anything about the magnetization pattern given the measured stray field, we must measure several materials parameters (section S4) and make further assumptions based on micromagnetic theory. Recent studies dealing with these issues have taken two main approaches: assume a domain wall profile based on micromagnetic theory \cite{Tetienne2015,Gross2016}, or fix a local magnetization gauge or helicity angle $\psi_h$ \cite{Dovzhenko2018}. The first method has the advantage that the local helicity angle can be extracted and need not be assumed as fixed along the entire length of a domain wall, but the second method has the nice property that it does not rely on an analytic form of the domain wall profile (DMI will lead to deviations in domain wall shape \cite{Thiaville2012}). In this work we use the reconstruction methods described in \cite{Dovzhenko2018} to estimate the domain wall position in the Bloch magnetization gauge ($\psi_h = 0$), but then fix the magnetization pattern using the analytic form of a thin film domain wall to calculate the stray magnetic field as a function of helicity angle. 

For the low external fields used in these measurements, the stray field along the NV axis changes sign at different points in the imaging plane. Since the NV measures only the absolute value of the stray field, estimation of the domain wall position requires us to assign a polarity to regions of the image separated by zero-field contours. Reconstruction of the full vector magnetic field can then proceed as described in \cite{Dovzhenko2018,Lima2009}. The $M_z$ component of magnetization is easily calculated in the Bloch gauge by projecting Fourier components of the stray field $\tilde{B}_z (\mathbf{k}, z)$ down to the sample surface using a stray field transfer function $\alpha(\mathbf{k},z,t)$, similar to \cite{Dreyer2007},
\begin{align}
\alpha(\mathbf{k},z,t)\equiv \frac{1}{2} e^{-kz}\left( 1-e^{-kt} \right) \\
\tilde{M}_z(\mathbf{k}) = \frac{\tilde{B}_z(\mathbf{k},z)}{k\alpha(\mathbf{k},z,t)}
\end{align}

\begin{figure}[h]
\includegraphics[width=\columnwidth]{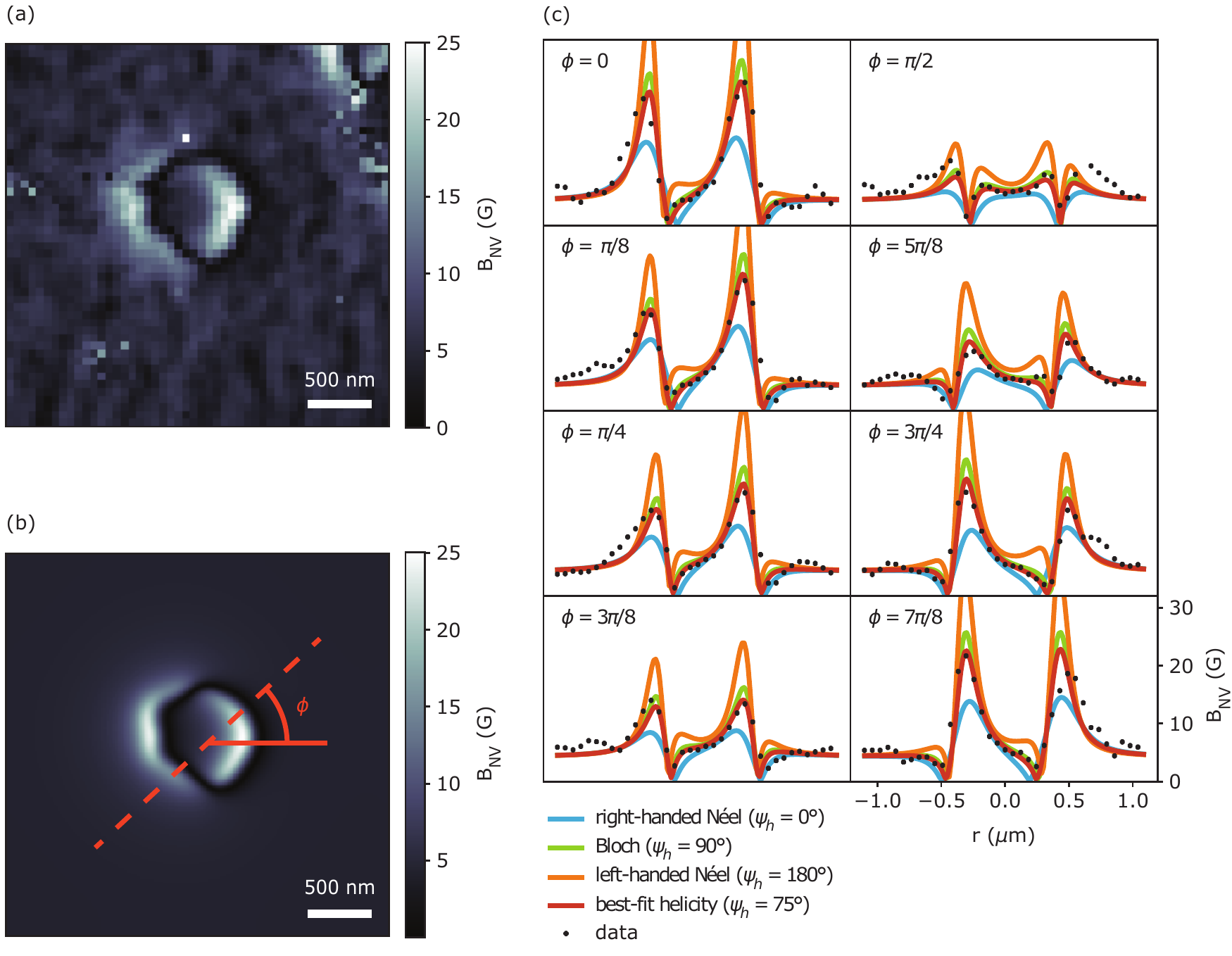}
\let\nobreakspace\relax
\caption{Magnetization reconstruction example. (a) Measured absolute value of the magnetic field along the NV axis. (b) The simulated magnetic field along the NV axis for the best fit helicity angle $72^{\circ}$. (c) Linecuts at various angles $\phi$ across the magnetic bubble, comparing the measured magnetic field shown in (a) to the simulated field for four domain wall types--- right-handed N\'eel, Bloch, left-handed N\'eel, and a domain wall with $\psi_h=75^{\circ}$. }
\label{mr1752}
\end{figure}

\begin{figure}[h]
\includegraphics[width=\columnwidth]{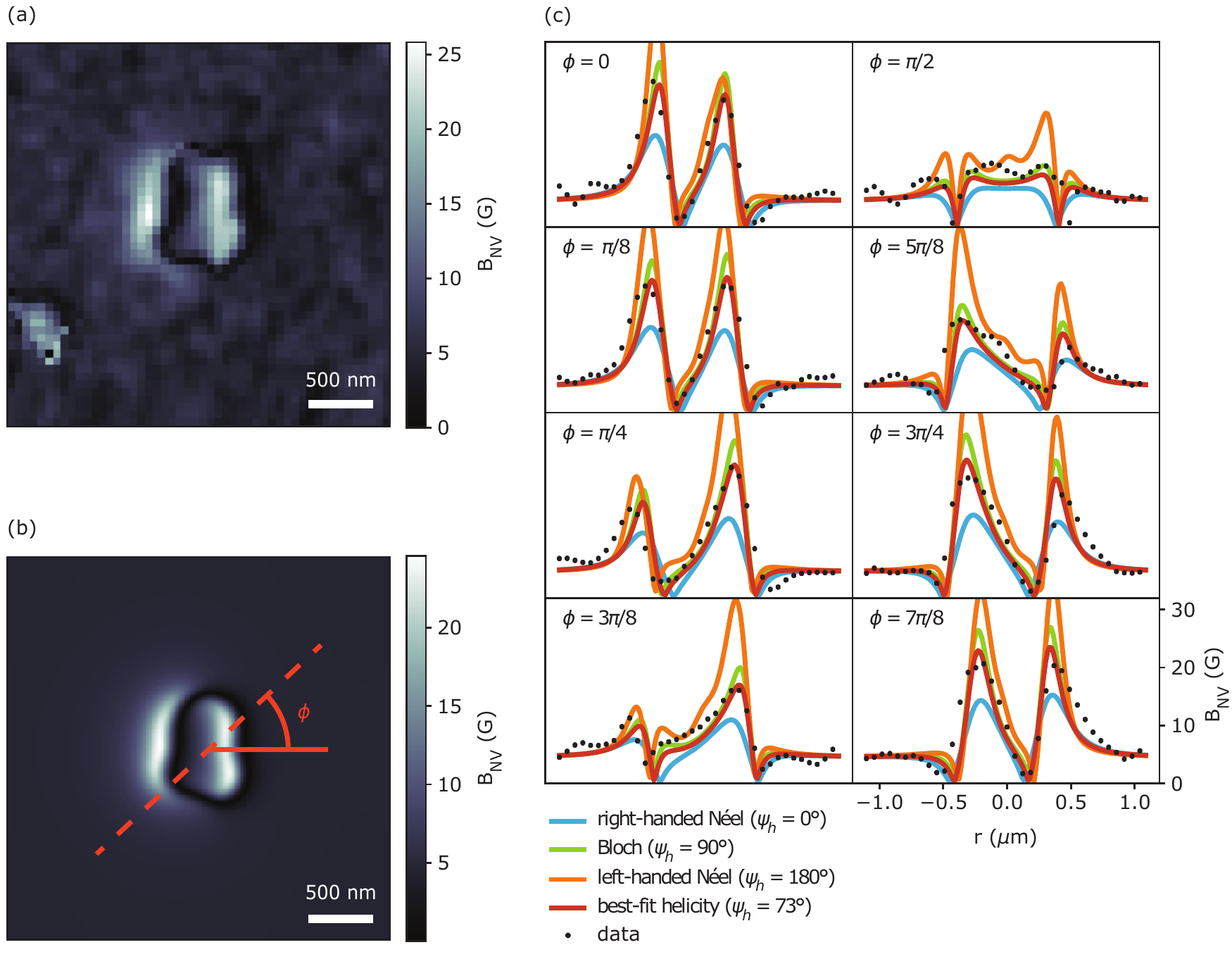}
\let\nobreakspace\relax
\caption{Magnetization reconstruction example. (a) Measured absolute value of the magnetic field along the NV axis. (b) The simulated magnetic field along the NV axis for the best fit helicity angle $69^{\circ}$. (c) Linecuts at various angles $\phi$ across the magnetic bubble, comparing the measured magnetic field shown in (a) to the simulated field for four domain wall types--- right-handed N\'eel, Bloch, left-handed N\'eel, and a domain wall with $\psi_h=73^{\circ}$. }
\label{mr1739}
\end{figure}

In our case, this procedure leads to amplification of image noise because of the somewhat large NV scan height. $M_z$ is then used to find the position of the domain wall, but is not used to simulate the stray field. The magnetization components calculated relative to the domain wall position are
\begin{align}
M_z = M_s \text{tanh}\left( \frac{x_{\perp}}{\Delta_{DW}} \right) \\
M_{\perp} = M_s \text{cos}(\psi_h) \text{sech}\left( \frac{x_{\perp}}{\Delta_{DW}} \right) \\
M_{\parallel} = M_s \text{sin}(\psi_h) \text{sech}\left( \frac{x_{\perp}}{\Delta_{DW}} \right)
\end{align}
\begin{align}
\psi_h =
\begin{cases}
0, & D>D_c \\
\text{arccos}\left( \frac{D}{D_c} \right), & |D|<D_c \\
\pi, & D<-D_c
\end{cases}
\end{align}\\
where $D_c = 2 \text{ln}(2)\mu_0 M_s^2 t/\pi^2$. Using the BLS value of $D$, the expected helicity angle is $\psi_h \simeq 53^{\circ}$. However, assuming a fixed helicity angle, the best fit helicity angles given by the images in Figure 4, S\ref{mr1752}, and S\ref{mr1739} are 66$^{\circ}$, 75$^{\circ}$, and 73$^{\circ}$ respectively. These best fit helicity values depend on the extracted material parameters described in section S4, and as such, the main source of uncertainty in determining these values is due to a possible systematic error caused by an unknown magnetic dead layer thickness. Using the the material parameters extracted for the case of the 0.36 nm thick magnetic dead layer described in section S4, the best fit helicity values for the skyrmion bubbles in Figure 4, S\ref{mr1752}, and S\ref{mr1739} are 80$^{\circ}$, 86$^{\circ}$, and 85$^{\circ}$ respectively.

\end{document}